\documentclass[12pt]{article}

\usepackage{times}
\usepackage{color,soul}

\usepackage{graphicx}

\newenvironment{sciabstract}{%
\begin{quote} \bf}
{\end{quote}}

\title{Discovery of an Unconventional Quantum Echo by Interference of Higgs Coherence 
}

\author
{C. Huang$^{1,2}$, M. Mootz$^{2}$, L. Luo$^{2}$, D. Cheng$^{2}$, J. M. Park$^{2}$, R. H. J. Kim$^{2}$, Y. Qiang$^{1,2}$, \\V. L. Quito$^{2}$, Yongxin Yao $^{2}$, P. P. Orth$^{1,2}$, I.~E.~Perakis$^{3}$ and J. Wang$^{1,2\dag} 
	$
	\\
	\normalsize{$^{1}$Department of Physics and Astronomy, Iowa State University, Ames, IA 50011 USA}\\
	\normalsize{$^{2}$Ames National Laboratory, Ames, IA 50011 USA}\\
	\normalsize{$^{3}$Department of Physics, University of Alabama at Birmingham,}\\
	\normalsize{Birmingham, AL 35294-1170, USA}\\
	\normalsize{$^\dag$To whom correspondence should be addressed; jgwang@iastate.edu, jgwang@ameslab.gov}
}

\date{}

\begin{document} 
	
	
	\baselineskip24pt
	
	
	\maketitle

	
	\begin{sciabstract}
Nonlinearities in quantum systems are fundamentally characterized by the interplay of phase coherences, their interference, and state transition amplitudes. 
Yet the question of how quantum coherence and interference manifest in transient, massive Higgs excitations, prevalent within both the quantum vacuum and superconductors, remains elusive. One hallmark example is photon echo, enabled by the generation, preservation, and retrieval of phase coherences amid multiple excitations. 
Here we reveal an unconventional quantum echo arising from the Higgs coherence in superconductors, and identify distinctive signatures attributed to Higgs anharmonicity. 
A terahertz pulse-pair modulation of the superconducting gap generates a ``time grating" of coherent Higgs population, which scatters echo signals distinct from conventional spin- and photon-echoes in atoms and semiconductors.
These manifestations appear as Higgs echo spectral peaks occurring at frequencies forbidden by equilibrium particle-hole symmetry, an ``asymmetric” delay in the echo formation from the dynamics of the ``reactive" superconducting state, and negative time signals arising from Higgs-quasiparticle anharmonic coupling. 
The Higgs interference and anharmonicity control the decoherence of driven superconductivity and may enable applications in 
quantum memory and entanglement.  
	\end{sciabstract}

When exposed to terahertz (THz) coherent driving, superconductors (SCs) manifest reactive behaviors rather than adiabatic responses~\cite{matsunaga2014, Matsunaga:2013,yang2018,yang2019lightwave,vaswani2019discovery,Giorgianni2019,hybrid-higgs, NatPhys,liu2023probing,katsumi2023revealing,Zhang2023,Buzzi2021,cheng2023evidence}. The 2$\Delta_\mathbf{SC}$ energy gap determined by the order parameter is not rigid, leading to the emergence of Higgs collective modes at $\omega_\mathbf{Higgs}\simeq$2$\Delta_\mathbf{SC}$.
The anharmonicity within the SC state and the evolution of its ``soft" low-energy gaps pose fundamental questions that diverge from the conventional framework applied to comprehend quantum echo phenomena.
First, in the observed spin/photon echo signals in semiconductors~\cite{Chemla,Wegener1990,Nelson,Kuehn2011} and atoms~\cite{Tian2003}, correlations within many-electron ground states generally have minimal impact, as the high-energy gap due to bandstructure maintains a rigid state through rapid adjustments to the comparatively slower photoexcitation.
In contrast, the energy gap 2$\Delta_\mathbf{SC}$, determined by the SC order parameter, changes during THz oscillation cycles with $\omega_\mathbf{0}\simeq\Delta_\mathbf{SC}$. This evolution makes the SC state evolve into a stationary non-equilibrium state distinct from any thermal ground state. 
Second, superconductivity is distinct as it presents both Higgs excitations featuring broken symmetry and quasi-particle (QP) excitations arising from pair breaking. However, unlike conventional bound states, discerning between the $\omega_\mathbf{Higgs}$ mode and the QP gap 2$\Delta_\mathbf{SC}$ is challenging 
since their excitation energies are in close proximity. 
Observing quantum echoes will reveal the robust phase coherence and interference effects, allowing for disentangling between Higgs and QPs. 
Additionally, the observation of echoes at $\omega_\mathbf{Higgs}$ is absent so far, and QP fluctuations mostly overshadow nonlinear spectra in the absence of inversion symmetry breaking or disorder~\cite{Klein1980,Littlewood1981,Murotani,yang2019lightwave,vaswani2019discovery,Shimano2019,katsumi2023revealing}.

Recently, notable progress has been made in coherent control schemes, particularly with THz-driven supercurrent by pulse-pair excitation that activates exotic collective excitations, including the hybrid-Higgs mode~\cite{hybrid-higgs} and phase-amplitude mode~\cite{NatPhys}.
Our work below explores the quantum echo process driven by the THz pulse-pair scheme. Figure~1A illustrates a three-step process that involves both Higgs and QP excitations. First (\#1), 
the nearly resonant coupling of the light field of pulse A (red wiggled arrows) via vector potential $A$ to the SC state, where $\omega_\mathbf{0}\simeq$2$\Delta_\mathbf{SC}$, induces a Higgs field excitation $\mathbf{H}(t)$ at time $t=0$ (blue dashed line). This excitation is facilitated via inversion-symmetry breaking \cite{moor, yang2019lightwave,vaswani2019discovery,Shimano2019} or disorder \cite{matsunaga2014}. 
We model the former process using the light-induced superfluid momentum  $\mathbf{p}_\mathrm{S}(t)$ (red straight arrow) discussed later.
This Higgs excitation subsequently dephases due to destructive interference with the QP continuum (Landau damping).
After the inter-pulse delay time $\tau>0$, the second light field, B, excites a second Higgs coherence, stops the above dephasing and stores the coherence in the order parameter amplitude (\#2) by creating a ``temporal grating" of coherent Higgs population along the time $\tau$ axis. This difference-frequency Raman process is controlled by the relative phase between A and B, which is tuned via the pulse time delay. 
A third field B induces a QP coherence also at time $t=\tau$, which scatters off the coherent Higgs grating and creates the quantum echo signal. The ABB echo described above manifests as nonlinear current $\mathbf{J}_{NL}$
with a delayed formation at time $t_\mathrm{c}$ from constructive interference.
Moreover, a negative time signal, referred to as BBA pulse sequence, 
is not expected for harmonic Higgs excitations. 
As discussed in supplementary Section 4, in this case, pulse B comes first and creates both a Higgs and a QP coherence. These evolve for inter-pulse delay $\tau$, at which time pulse A can sense them. 
Therefore, the negative time signals observed show storage of the 2-excitation coherence within the time-evolving quantum system due to anharmonicity arising from Higgs-QP couplings.

\begin{figure}[tbp]
	\begin{center}
		\includegraphics[width=150mm]{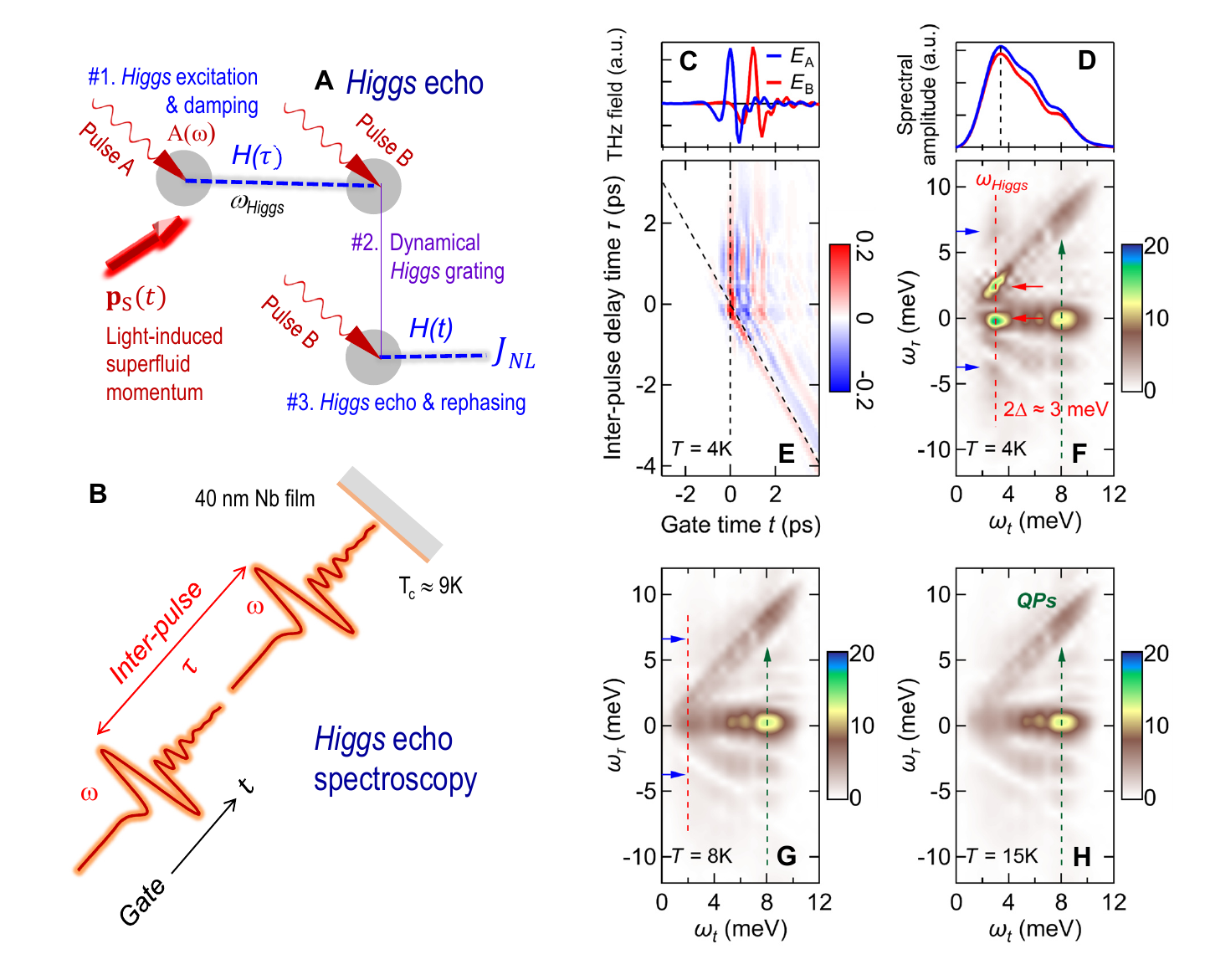}
	\end{center}
	\noindent {\textbf{Figure 1. Higgs echo spectroscopy in a niobium superconductor through excitation by THz pulse pairs.} 
		(\textbf{A}) Schematics of the quantum echo process as a three-step process that involves quantum interference of both Higgs (\#1 and \#2) and coherent QP (\#3) excitations (see main text). The response of SC condensate during cycles of THz electric fields (red wiggle line) gives rise to momentum $\mathbf{p}_\mathrm{S}(t)$ (red arrow) that breaks the inversion symmetry and enables the dipole-forbidden, Higgs excitations.
		(\textbf{B}) Schematics of THz multidimensional coherent spectroscopy of a 40~nm Nb film. Two delay lines are employed to control THz-pulse-pair excitation, both the inter-pulse delay $\tau$ and the sampling delay $t$. The nonlinear signals are collected using a double modulation scheme (Supplementary Section 3). 
	(\textbf{C}) Temporal waveforms of the nearly single-cycle THz pulse--pair following transmission through the Nb film (red and blue lines), and (\textbf{D}), spectra of the used pulses centered at 
		$\omega_0\sim$4~meV (vertical dashed line) and with broadband frequency width $\Delta\omega\sim$6~meV.
		(\textbf{E}) Two-dimensional (2D) false-colour plot of the measured coherent nonlinear transmission $E_\mathrm{NL}(t,\tau)$ in our sample induced by THz pump electric fields of $E_\mathrm{THz, A, B}\sim$50\,kV cm$^{-1}$ at 4K. (\textbf{F}) The Fourier transform is applied to the $E_\mathrm{NL}(t,\tau)$ trace depicted in panel (E).
		(\textbf{G}) and (\textbf{H}) 2D Fourier transform of $E_\mathrm{NL}(t,\tau)$ employing the identical THz pump field conditions as (\textbf{E}) for temperatures of 8~K and 15~K. Note that the subtle yet discernible spectral difference between Figs. 1G and 1H is depicted in Fig. 2B. }
\end{figure}

In this report, we demonstrate quantum echoes involving transient, yet enduring, Higgs coherence. We address two significant questions: (1) How to induce the wave-mixing process sketched in Fig.~1A given that nonlinear currents $\mathbf{J}_{NL}$ from Higgs excitations is precluded by the symmetry in the ground state? 
(2) What are the unique characteristics of Higgs echoes in a ``reactive” SC state that distinguish them from the conventional photon echoes?  
We implement THz Higgs echo spectroscopy 
in a niobium (Nb) superconductor as illustrated in Fig.~1B. We follow a phase-resolved and collinear, pulse-pair excitation protocol within the framework of multi-dimensional coherent spectroscopy (MDCS)~\cite{mukamel,katsumi2023revealing,NatPhys,Kuehn2011,Somma2014,Junginger2012,maag2016,Nelson,Johnson2019,Pal2021,Tarekegne2020,Mahmood, NatPhys,Blank2023}. 
Unlike for the conventional pump-probe protocol, which has been used to access a single Higgs or QP excitation pathway~\cite{Matsunaga:2013,matsunaga2014,Matsunaga2017,hybrid-higgs,Chu2020}, the MDCS scheme employed here enables the observation of photon echo signals arising from the interference of multiple phase-coherent excitations. 
Taking advantage of both the real time $t$ and the relative phase delay time $\tau$ between the two THz fields, we can separate Higgs echo peaks from other nonlinear pathways in a 2D frequency plane defined by the excitation $\omega_t$ and detection $\omega_{\tau}$ frequencies. 
In this manner, we discern distinct and sharp Higgs echo peaks in MDCS spectra at low temperatures and low fluences. We observe a delayed echo formation influenced by the time-dependent SC bands and supercurrent flow, accompanied by a pronounced negative time signal associated with Higgs anharmonicity.
These features are well reproduced by our quantum kinetic simulations, pinpointing the salient roles of the coherent Higgs grating and Higgs-QP coherent coupling. 

We study a Nb film of 40nm thickness on a 1mm Si substrate (Supplementary section~1). The sample exhibits a SC transition at T$_\mathrm{C} \sim$9\,~K and displays a SC gap of 2$\Delta_\mathbf{SC}\sim$~3.1~meV as characterized by equilibrium THz conductivity measurements (Supplementary Section~2). Our THz-MDCS 
(Supplementary section~3) measures all detected coherent nonlinear responses to two phase-locked, nearly single-cycle THz pulses, A and B, of similar field strengths (Figs.~1C). The two pulses have central frequency $\omega_0\sim$4~meV above the SC energy gap (black dashed line, Fig.~1D). The broadband pulse frequency width, $\Delta\omega_0\sim$~6~meV, excites a wide range of the QP continuum, in addition to the Higgs mode. 
The measured phase-resolved, nonlinear supercurrent differential emission, $E_\mathrm{NL}(t,\tau)=E_\mathrm{AB}(t,\tau)-E_\mathrm{A}(t)-E_\mathrm{B}(t,\tau)$, is  recorded as a function of both the gate time $t$ and the delay time $\tau=t_\mathrm{A}-t_\mathrm{B}$ between the two pulses A and B (red double-arrow, Fig. 1B). 
The $\tau$ axis facilitates the generation of a {\em temporal} grating, which scatters $N$-wave mixing signals into different $\omega_{\tau}$ components. The process is analogous to the traditional spatial grating which scatters wave-mixing signals along various directions~\cite{Cundiff, mukamel}. 

A representative 2D temporal profile of the THz coherent nonlinear dynamics of $E_\mathrm{NL}(t,\tau)$ driven by low fields $E_\mathrm{THz, A, B}\sim$50\,kV cm$^{-1}$ is shown in Fig.~1E. 
This figure reveals pronounced coherent temporal $\tau$-oscillations that last much longer than the temporal overlap between the two driving pulses (Fig.~1C vs Fig.~1E), which indicates the presence of long-lived coherent excitations. 
We underpin the above coherent responses by performing Fourier transformations of $E_\mathrm{NL}(t,\tau)$ with respect to both $t$ (frequency $\omega_t$) and $\tau$ (frequency $\omega_\tau$). 
Figures~1F-1H compare these experimentally-measured 2D spectra (unnormalized) at 4~K, 8~K, and 15~K across the superconducting transition temperature of 9~K. We observe multiple sharp peaks with the same $\omega_t=\omega_\mathrm{H} \sim$2$\Delta_\mathrm{SC} \simeq$3~meV (red dashed line) that split along the vertical $\omega_\tau$-axis.
These peaks change drastically as the temperature increases above  $T_\mathrm{C}$. Additional peaks are observed along $\omega_t \simeq 2\omega_\mathbf{0}\sim$8~meV, indicated by the green arrows in Figs.~1F-1H.
The latter peaks persist both below and above $T_\mathrm{C}$, which aligns with our expectations for QP excitations. 

We highlight two key points. First, the observed peaks (red arrows in Fig.~1F for $T=4$~K)
along the $\omega_t\simeq$2$\Delta_\mathrm{SC}=\omega_\mathrm{H}$ red dashed vertical line have frequency widths $\sim$0.5~meV, i.~e., more than one order of magnitude smaller than the pulse bandwidth $\Delta\omega_0$ (Fig.~1D). 
This narrow linewidth indicates that, below $T_\mathrm{C}$, the coherent time evolution indicated by $E_\mathrm{NL}(t,\tau)$ is governed by long-lived oscillations. 
The observed MDCS peaks at $\omega_t=\omega_\mathrm{H}$ exhibit a magnitude surpassing that of conventional nonlinear wave-mixing signals at $\omega_t \simeq n \omega_\mathbf{0}$ ($n=1,2...$), showcasing a unique feature consistent with our anticipated coherent response from a ``reactive" ground state. Remarkably, MDCS peaks with $\omega_t=\omega_\mathrm{H}$ are not allowed by the equilibrium SC state symmetry~\cite{Mootz2022, yang2019lightwave}. Instead, they can arise from the Higgs collective modes of a non-equilibrium SC state with light-induced inversion symmetry breaking and/or disorder. The state corresponds to finite-momentum Cooper pairing that evolves during the pulse but lives well after it~\cite{Mootz2022, NatPhys}. Furthermore, the sharp $\omega_t=\omega_\mathrm{Higgs}$ peaks exhibit significant broadening, redshift, and reduction as the temperature approaches $T_\mathrm{C}$. They vanish above $T_\mathrm{C}$. 
This becomes apparent when examining the spectra at 4 K (Fig. 1F), denoted by the red dashed lines, in contrast to those at 8 K (Fig. 1G) and 15 K (Fig. 1H).

\begin{figure}[tbp]
	\begin{center}
		\includegraphics[width=155mm]{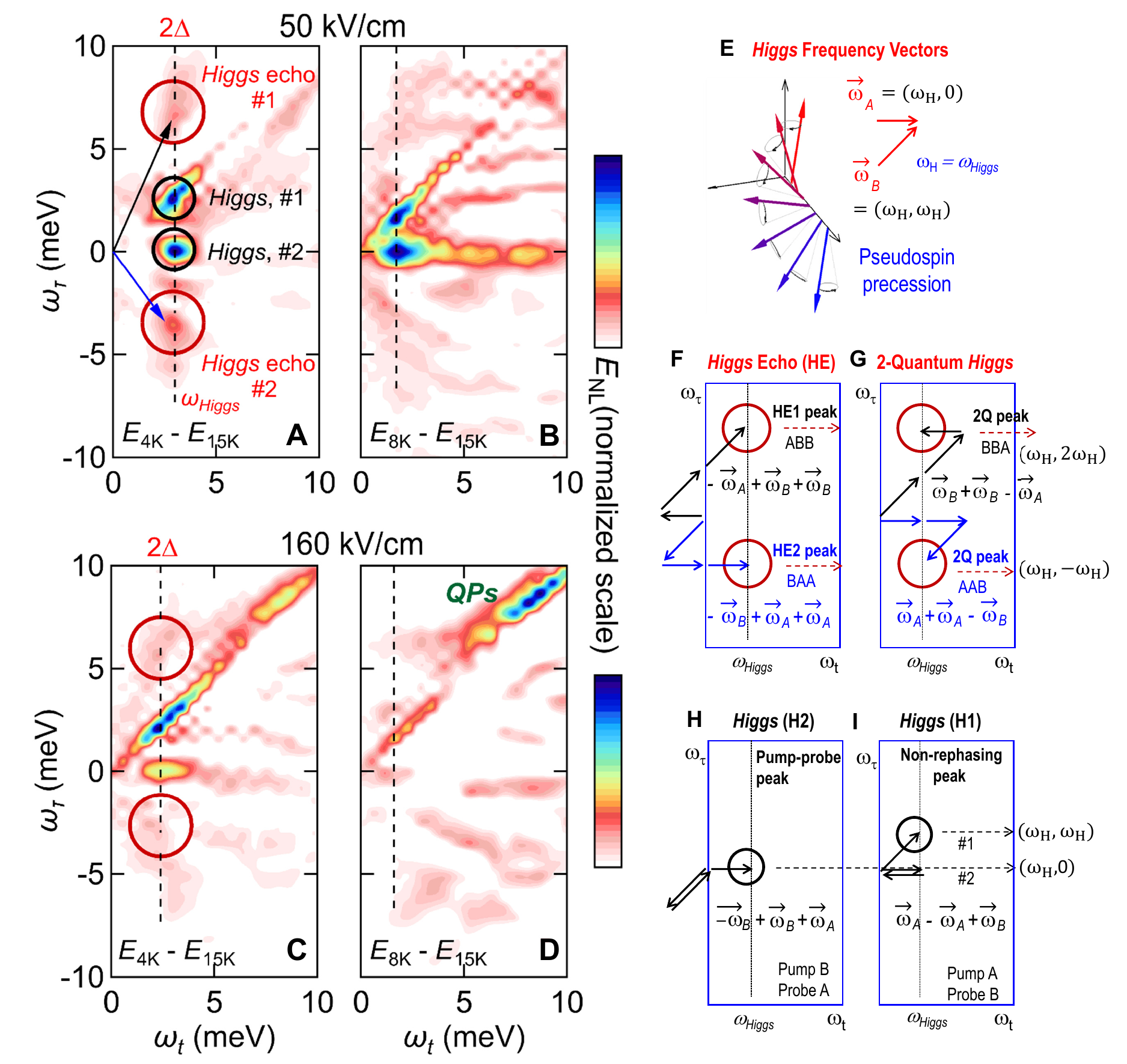}
	\end{center}
	\noindent {\textbf{Figure 2. Variations of Higgs echo and other collective excitation peaks in MDCS spectra by changing the THz driving field and temperature.} 
		The normalized THz-MDCS spectra within superconducting states are obtained through the subtraction of normal state data for both low and high driving fields. \textbf{A}: $E_\mathrm{THz, A, B}\sim$50\,kV cm$^{-1}$ and 4~K; \textbf{B}: $\sim$50\,kV cm$^{-1}$ and 8~K; \textbf{C}: $\sim$160\,kV cm$^{-1}$ and 4~K; \textbf{D}: $\sim$160\,kV cm$^{-1}$ and 8~K.
		\textbf{E} The Higgs frequency vectors, denoted as $\omega_\mathrm{A}=(\omega_\mathrm{H},0)$ and $\omega_\mathrm{B}=(\omega_\mathrm{H},\omega_\mathrm{H})$, 	illustrate pseudo-spin coherent oscillations induced by THz-pulse pair excitations.	
\textbf{F} Schematics of the quantum dynamical pathways for generating two Higgs echo peaks employing ABB (HE1) and BAA (HE2) pulse sequences. 
Other collective excitation peaks in MDCS spectra are depicted as follows: \textbf{G}, 2-quantum Higgs peaks; \textbf{H}, Higgs (H2) peak by pump-probe process; \textbf{I}: Higgs (H1) peak by non-rephasing process. }
\end{figure}
	 	
Second, the THz-MDCS spectrum at 4~K (Fig~1F) exhibits four distinct peaks at $\omega_t=\omega_\mathrm{H}$. In addition to the two main Higgs peaks (the red arrows) discussed above, two additional peaks are seen along the vertical $\omega_{\tau}$ axis  (the blue arrows). 
The suppression of these additional $\omega_{\tau}$ components as we approach $T_\mathrm{C}$ is evident when comparing the spectra at 4~K (Fig.~1F) and 8~K (Fig.~1G).
They come from the Higgs echo process illustrated in Fig.~1A as elaborated below.  
To better visualize these subtle, yet prominent, peaks,  
we subtract the spectrum acquired at 15~K from the experimental $E_\mathrm{NL}(\omega_t,\omega_\tau)$ spectrum recorded below $T_\mathrm{C}$ to remove any temperature-independent spectral artifacts. 
We present the normalized spectra obtained after this subtraction in Figs.~2A-2D for both low and high driving fields, at $T=4$~K and $T=8$~K. 
We focus on Fig.~2A obtained at $E_0=50$~kV cm$^{-1}$ and low temperature of 4~K. The $E_\mathrm{NL}(\omega_t,\omega_\tau)$ spectrum clearly reveals the presence of the four prominent peaks along the vertical dashed line positioned at $\omega_t\simeq 2\Delta_\mathrm{SC}$.
Most intriguingly, apart from the stronger peaks (black circles) at roughly $(\omega_\mathrm{H},0)$ (H2) and $(\omega_\mathrm{H},\omega_\mathrm{H})$ (H1), 
the two Higgs echo peaks distinctly emerge at $(\omega_\mathrm{H},2\omega_\mathrm{H})$ (HE1) and $(\omega_\mathrm{H},-\omega_\mathrm{H})$ (HE2) (red circles), as confirmed by our analysis and simulations below.  
	
The Higgs echo features become prominent only when Higgs collective excitations are long-lived, with sharp Higgs peaks dominating the MDCS spectrum, indicating 
that dissipation is suppressed.	
When either the pump field is increased to $E_0=160$~kV cm$^{-1}$ (Fig.~ 2C) or the temperature is raised to 8~K (Fig.~2B), a substantial reduction and broadening in the intensity of the HE1 and HE2 Higgs echo peaks are seen, as indicated within the two red circles in Fig.~2C. 
This observation coincides with the broadening and red-shift of the two main Higgs peaks, H1 and H2, consistent with either excitation or temperature-induced dephasing, discussed later.
For strong pump field  $E_0=160$~kV cm$^{-1}$ at 8K (Fig.~2D), the primary features in the $E_\mathrm{NL}(\omega_t,\omega_\tau)$ spectrum are the diagonal QP excitation peaks located roughly at $(2\omega_\mathrm{0},2\omega_\mathrm{0})$ 
(Figs.~2C and 2D).

The prevalence of the peaks H1 and H2, centered around frequencies $\omega_\mathrm{H} < \omega_0$ and $2 \omega_0$, suggests analyzing the MDCS spectra by introducing Higgs frequency vectors that describe collective modes of the light-induced non-equilibrium SC state: $\omega_\mathrm{A}=(\omega_\mathrm{H},0)$ and $\omega_\mathrm{B}=(\omega_\mathrm{H},\omega_\mathrm{H})$, respectively, as depicted in Fig. 2E. 
Analogous to the formation of a spatial grating, 
the $\omega_\mathrm{A,B}$ frequency vectors allow us to describe the formation of a temporal, coherent Higgs grating along the $\omega_{\tau}$ axis via the difference-frequency Raman processes $\pm (\omega_\mathrm{A} - \omega_\mathrm{B})$.
Figure~2F illustrates the nonlinear quantum excitation processes $ 2 \omega_\mathrm{B} - \omega_\mathrm{A}$ and $2 \omega_\mathrm{A} - \omega_\mathrm{B}$ that give rise to Higgs echo peaks HE1 and HE2. 
In particular, $(-\omega_\mathrm{A}+\omega_\mathrm{B})  + \omega_\mathrm{B}$ (ABB echo, black arrows) and $(-\omega_\mathrm{B}+\omega_\mathrm{A}) + \omega_\mathrm{A}$ (BAA echo, blue arrows) describe the scattering of a SC excitation $\omega_\mathrm{B}$ ($\omega_\mathrm{A}$) off the temporal grating created by $\omega_\mathrm{B}-\omega_\mathrm{A}$ ($\omega_\mathrm{A}-\omega_\mathrm{B}$), leading to the HE1 and HE2 peaks 
at positions $(\omega_\mathrm{H},2\omega_\mathrm{H})$ and $(\omega_\mathrm{H},-\omega_\mathrm{H})$. 
The negative time signal for the ABB (BAA) echo manifest as a 2-quantum Higgs peak at the same position as HE1 (HE2), as illustrated in Fig. ~2G (2Q peaks), which as we discuss later can arise from Higgs-QP coherent coupling.
In contrast, the H1(H2) Higgs peak at $\omega_\mathrm{B}=(\omega_\mathrm{H},\omega_\mathrm{H})$ ($(\omega_\mathrm{H},0)$) is generated by the pump-probe and non-rephasing nonlinear process $(\omega_\mathrm{A} - \omega_\mathrm{A}) + \omega_\mathrm{B}$ ($(-\omega_\mathrm{B}+\omega_\mathrm{B}) + \omega_\mathrm{A}$) in Figs.~2H and 2I. 

To differentiate the temporal evolution and quantum pathway in more detail, we apply a 2D-Gaussian spectral filter to isolate the HE1 and HE2 peaks indicated by the red circles in Fig.~2A, and then transform the echo signals back into the time domain. The signals extracted from the ABB and BAA Higgs-echo peaks in this way are presented in Figs.~3A and 3B, respectively. The order in which the pulses interact with the superconductor determines the well-defined phase fronts of these Higgs-echo signals. Specifically, for the ABB pulse sequence, the phase fronts are oriented parallel to the blue dashed line in Fig.~3A. 
On the other hand, employing the BAA pulse sequence results in phase fronts that are nearly perpendicular to the blue dashed line in Fig.~3B. 
Importantly, the ABB (BAA) signal is nonzero even when pulse B (A) arrives before pulse A (B). This time ordering, opposite to the one shown in Fig.~1A, gives the negative time signals that corresponds to 
the BBA (AAB) sequence marked in Fig.~3A (Fig.~3B).
The remarkable negative time signals observed here are exceptional and suggest a 2-quantum excitation and anharmonicity. 
Our simulations identify the Higgs and QP coupling as the origin.  

\begin{figure}[tbp]
	\begin{center}
		\includegraphics[width=150mm]{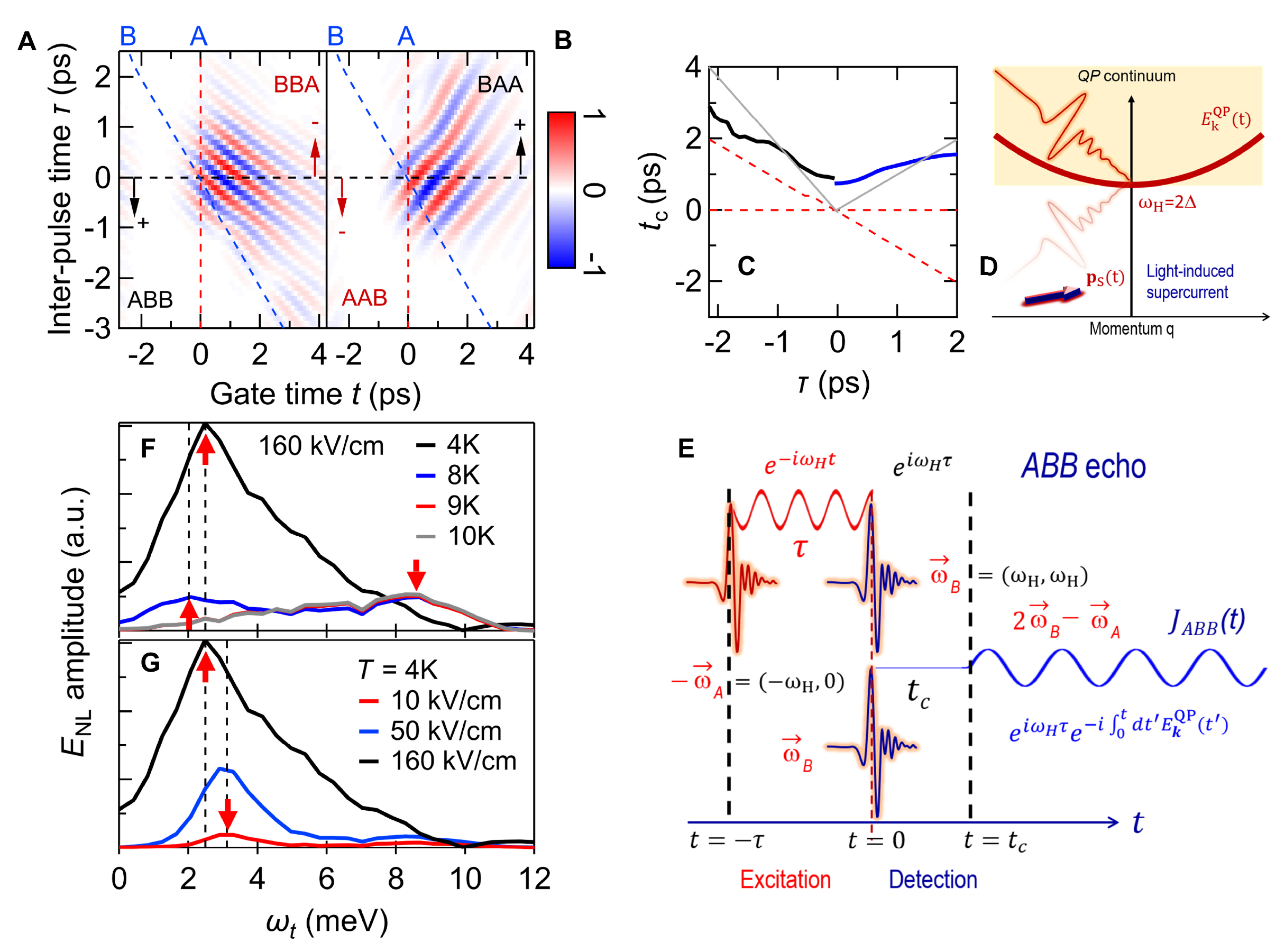}
	\end{center}
	\noindent 	{\textbf{Figure 3. Identification and origin of the Higgs echo and 2-Higgs quantum peaks in THz-MDCS signals.} 
		(\textbf{A} and \textbf{B}) Higgs-echo signals ABB and BAA in time domain resulting from the inverse Fourier transformation of the signals indicated by red circles in Fig.2A, consistent with the HE1 and HE2 echo signals at positive times from the processes illustrated in Fig.~2F. In addition, the ABB (BAA) pulse sequences also show a negative time signal (red arrows) when pulse B (A) arrives before pulse A (B), originating from the 2-Higgs quantum peaks illustrated in Fig.~2G. The positive (negative) time directions  are indicated by black arrows (red arrows). 
		(\textbf{C}) Temporal positions of the ABB and BAA echoes, $t_\mathrm{C}$ (blue and black lines), are time-delayed compared to both driving pulses (dashed red lines). Remarkably, the echo formation time exhibits a pronounced departure from symmetric temporal positions in relation to the pulse-pair excitations (depicted by gray lines), deviating notably from conventional photon echoes. 
		(\textbf{D} and \textbf{E}) An illustration of the ABB echo formation process (Fig.~1E) from the dynamical response of the ``reactive" superconducting state and supercurrent, as presented in Fig.~1D. $E_\mathbf{k}^\mathrm{QP}(t)$ denotes a time-dependent QP energy dispersion. 
		(\textbf{F}) The Fourier spectra of two-pulse THz coherent signals $E_\mathrm{NL}$ at various temperatures from 4~K to 10~K for a peak THz pump electric field of  $E_\mathrm{THz, A, B}\sim$160~kV/cm and $\tau=$3~ps. 
		(\textbf{G}) The Fourier spectra of $E_\mathrm{NL}$ at various pump fields at temperature of 4~K and $\tau=$3~ps.} 
\end{figure}	

To further quantify the unconventional nature of the quantum echo signals observed, we define the temporal position of any time-dependent signal $E(t)$ as $t_\mathrm{c}=\int\mathrm{d}t\,t E^2(t)/\int\mathrm{d}t\,E^2(t)$, i.e., the center of gravity~\cite{Kuehn2009}. 
Figure~3C shows this $t_\mathrm{c}$ for the Higgs echo signals HE1 and HE2 (black and blue solid lines) and compares it to that of the two individual laser pulses A and B (dashed red lines). The ABB echo signal primarily contributes at $\tau>0$ (blue line), while $t_\mathrm{c}$ at  $\tau<0$ 
(black line) is dominated by the BAA echo signal. 
Note that the Higgs echo is {\em time-delayed} as compared to both pulses. 
Such a behavior is the hallmark for the photon echoes expected from inhomogeneously broadened systems, as, with purely homogeneous broadening, $t_\mathrm{c}$ should coincide with the $t_\mathrm{c}$ of the last pulse (red dashed lines, Fig.~3C) due to a free-induction decay. 
Most notably, the delayed echo formation time markedly contrasts to the symmetric time $t_\mathrm{c}=\tau$ (gray lines, Fig.~3C) after the last pulse observed in conventional echoes~\cite{Wegener1990,Schmitt-Rink1990b,Koch1992}. 

The unique characteristics of Higgs echoes originate from the combined influence of the ``reactive" oscillating gap amplitude $|\Delta(t,\tau)|$ and the damping of the light-induced superfluid momentum $\mathbf{p}_\mathrm{S}(t)$. This interplay results in a {\em time-dependent} QP energy dispersion $E_\mathbf{k}^\mathrm{QP}(t)$ as illustrated in Fig.~3D. 
Excitation by a broadband pulse (Fig.~1D) leads to both Higgs modes and a broad inhomogeneous QP distribution with energy $E_\mathbf{k}^\mathrm{QP}(t)$. The latter results in quantum echo rather than free induction decay. Figure~3E illustrates the interference processes for the ABB echo. The phase accumulation at QP momentum $\mathbf{k}$ is $\omega_\mathrm{H}\tau -\int_{0}^{t} E_\mathbf{k}^\mathrm{QP}(t')\cdot dt'$. The first term accounts for the initial Higgs time evolution up to time $t=0$ (red wiggle lines). The second term describes the subsequent, re-phasing time evolution (blue wiggle lines) by coherent QP excitation. 
The time-dependence of $E_\mathbf{k}^\mathrm{QP}(t)$ involving both Higgs and QP coherent excitations results in the observed distinct ``asymmetric” delay in echo formation in Fig. 3C.

Figures~3F and 3G reveal pronounced temperature- and fluence-induced Higgs dephasing process.  
Figure~3F shows the temperature-dependent Fourier transform of the MDCS temporal ($t$) dynamics for 4~K--10~K at fixed $\tau=$3~ps.  
As we approach $T_\mathrm{C}$ from below,  this signal rapidly decreases. Simultaneously, the frequency of the main peak below $\omega_0$ undergoes a redshift, and eventually gives way to a QP peak centered around $2 \omega_0\sim$8~meV. 
This is seen by comparing the 4~K (black line) and 8~K (blue line) versus 10~K (red line) traces (black dashed lines), consistent with 
our experiment (Figs~1F--1H).
Figure~3G shows three representative Fourier spectra of the MDCS temporal dynamics at $T$=4\,~K, obtained 
at $E_{\mathrm{pump}}=$10~kV/cm (red), 50~kV/cm (blue) and 160~kV/cm (black). 
We observe a redshift and broadening of the Higgs mode, accompanied by the generation of QPs (black dashed lines). Thermal QP excitations suppress both Higgs and Higgs echo signals consistent with our data in Figs. 2A-2D.

\begin{figure}[tbp]
	\begin{center}
		\includegraphics[width=160mm]{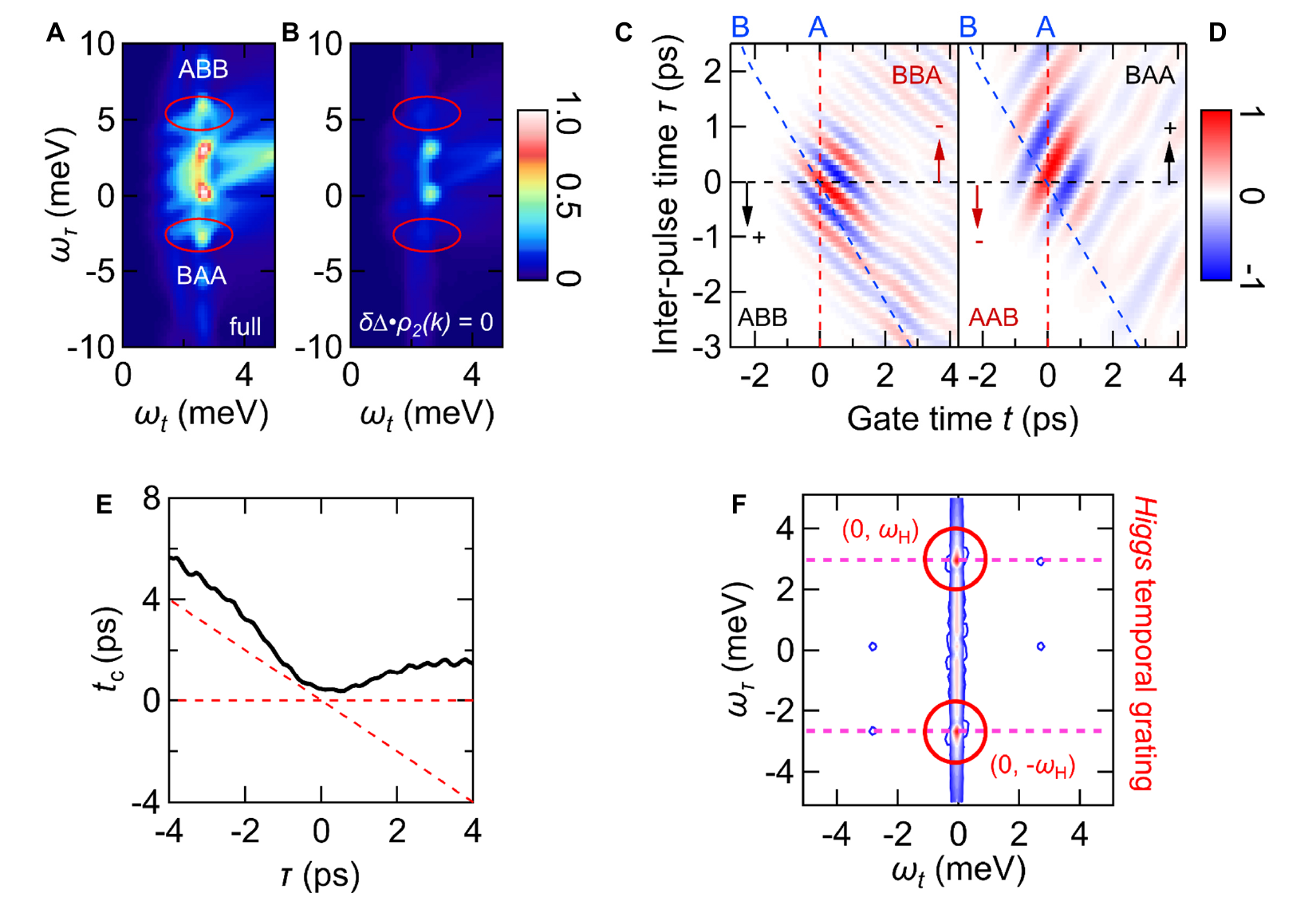}
	\end{center}
	\noindent {\textbf{Figure 4. Quantum kinetic simulation of Higgs echo signals from THz-MDCS spectra.}
		(\textbf{A}) Simulated nonlinear spectrum $E_\mathrm{NL}(\omega_t,\omega_\tau)$ , which accurately captures all of the dominant peaks observed in the measured nonlinear spectrum in Fig.~2A. The ABB and BAA Higgs echo signals are indicated by red circles.  (\textbf{B}) $E_\mathrm{NL}(\omega_t,\omega_\tau)$  for a calculation where $\delta|\Delta|\rho_2(k)=0$ in the simulations (Supplementary Section 5). The quantum echo signals BAA and ABB (red circles) are strongly suppressed in this case. (\textbf{C} and \textbf{D}) Higgs-echo signals ABB and BAA in time domain resulting from the inverse Fourier transformation of the signals indicated by circles in Fig.~4A. The ABB (BAA) pulse sequences shows a negative time signal when pulse B (A) arrives before pulse A (B), consistent with the experimental observations in Figs.~3A-3B. The positive (negative) time directions  are indicated by black arrows (red arrows). (\textbf{E}) Calculated temporal position of the photon echo, $t_\mathrm{c}$ (black line), which is time-delayed compared to both driving pulses (dashed red lines). (\textbf{F}) Two-dimensional spectrum of the order parameter modulation $\delta|\Delta|$ that creates the Higgs temporal grating. Horizontal dashed lines indicate $\omega_{\tau}=\pm\omega_\mathrm{H}$. The spectrum shows two strong peaks at $(0, \pm\omega_\mathrm{H})$ (red circles) which are generated by the difference frequency processes $\pm(\omega_\mathrm{A}-\omega_\mathrm{B})$ that require two time-delayed pulses with well-defined relative phase.}  
\end{figure}

To solidify the above physical depiction of unconventional Higgs quantum echoes, we conducted quantum kinetic simulations using the gauge-invariant superconductor Bloch equation theory~\cite{Mootz2020,Mootz2022,mootz2023multidimensional}.   
We directly simulated the nonlinear temporal dynamics measured in the experiment for a one-band $s$-wave BCS superconductor driven by two intense phase-locked terahertz electric-field pulses (Supplementary section~5). Figure~4A presents the simulated nonlinear response $E_\mathrm{NL}(t,\tau)$, which accurately captures all of the dominant peaks observed in the measured nonlinear spectrum in Fig.~2A. Similar to the experiment, the calculated 2D spectrum shows Higgs peaks at $(\omega_\mathbf{H},\omega_\mathbf{H})$ and $(\omega_\mathbf{H},0)$, while Higgs echo signals emerge at  $(\omega_\mathbf{H},-\omega_\mathbf{H})$ and $(\omega_\mathbf{H},2\omega_\mathbf{H})$. Analogous to the photon echo analysis presented in Fig.~3, we filter out the BAA and ABB Higgs-echo peaks in Fig.~4A and transform them back to the time domain. The resulting signals in the time domain are shown in Figs.~4C and 4D. The ABB (BAA) pulse sequences give a strong negative time signal even when pulse B (A) arrives before pulse A (B), and the phase fronts of both echoes are nearly perpendicular to each other. These results are fully consistent with our experimental observations in Figs.~3A-3B. 
The calculated temporal position of the photon echo, $t_\mathrm{c}$, is plotted in Fig.~4E (black line). It is time-delayed when compared to both driving pulses (dashed red lines) and deviates from the conventional symmetric echo formation time, in agreement with the experimental result in Fig.~3C. 

To clarify the origin of the echo signals in our calculations, we present in Fig.~4B the result of the simulation when setting $\delta|\Delta|=0$, 
where $\delta|\Delta|=(|\Delta_\mathrm{AB}(t,\tau)|- |\Delta_0|)-(|\Delta_\mathrm{A}(t)|- |\Delta_0|) – (|\Delta_\mathrm{B}(t,\tau)|- |\Delta_0|)$ and $|\Delta_0|$ denotes the equilibrium order parameter amplitude. 
$|\Delta_\mathrm{AB}(t,\tau)|$ describes the order parameter amplitude dynamics driven by coherent modulation of the order parameter amplitude by two THz pulses with well-defined relative phase controlled by $\tau$. 
Single pulse A (B) excitation results in $|\Delta_\mathrm{A}(t)|$ ($|\Delta_\mathrm{B}(t,\tau)|$). 
If pulses A and B act independently to create fluctuations of the order parameter from equilibrium, we have uncorrelated excitation with $\delta|\Delta|=0$.  
Clearly, the quantum echo signals HE1 and HE2 are strongly suppressed when we set $\delta|\Delta|=0$ in Fig.~4B.
Such an independent SC excitation by two THz fields is distinctly different from the pulse-pair driving 
result of $\delta|\Delta| \ne 0$ in Fig.~4A. 

We elaborate below that finite $\delta|\Delta|$ generates  
a dynamical Higgs grating and 
quantum echo signals.
Fig.~4A unravels the role of time-delayed coherent excitations with well defined relative phases and distinguishes it from the contributions of individual Higgs or QP excitations to the THz-MDCS  spectrum.   
As shown in the Supplement Section 5, we formulate the gauge invariant density matrix equations of motion in terms of four-component Anderson pseudo-spin oscillators $\rho_i$, $i=0,1,2,3$. 
The photon echo signals in our simulations are driven by the correlated contribution by the pulse-pair, $\delta\rho_2(t,\tau)=\Delta\rho^\mathrm{AB}_2(t,\tau)-\Delta\rho^\mathrm{A}_2(t)-\Delta\rho^\mathrm{B}_2(t,\tau)$. 
Here, $\Delta\rho^\mathrm{AB}_2(t,\tau)$ describes pseudo-spin canting away from the equilibrium $x-z$ plane by both phase-coherent pulses, while $\Delta\rho^\mathrm{A}_2(t)$ ($\Delta\rho^\mathrm{B}_2(t,\tau)$) describes pseudo-spin canting induced by pulse A (B) alone. 
By comparing the results of the full calculation in Fig.~4A, with the simulation in Fig.~4B, where $\delta|\Delta|= 0$, we show that the dominant contribution to the echo peaks originates from parametric driving by the $\delta|\Delta|$ oscillations in time $\tau$, which leads to a nonlinear driving force of the form $\delta|\Delta(t,\tau)|\Delta\rho^\mathrm{A,B}_2(t)$ that is additional to the conventional Raman  processes (discussed further in the Supplement Section 5). In particular, the 2D spectrum of $\delta|\Delta|$ in Fig.~ 4F shows two different peaks at $\omega_t=0$, which are displaced along the vertical $\omega_\tau$ axis that measures the phase coherence. These peaks of $\delta|\Delta|$, located at $(0,\pm\omega_\mathrm{H})$, are generated by  difference-frequency processes $\pm(\omega_\mathrm{A}-\omega_\mathrm{B})$. 
$\delta|\Delta(t,\tau)|$ thus describes a temporal grating that scatters a QP coherence $|\Delta\rho^\mathrm{A,B}_2(t)|$, to be contrasted with order parameter quench with $\tau \sim 0$. The grating is created by Higgs excitations A and B with well-defined relative phase.  At the same time, $\Delta\rho^\mathrm{A,B}_2(t)$ describes the continuum of photoexcited QP coherences with different ${\bf k}$. The time-delayed quantum echo signal can be understood in terms of refraction of such QP excitations from the dynamic Higgs grating. 
This process is comparable to the photon echo generation process in semiconductors, where the refracted signal scatters off a coherent population grating generated by the interference of two polarizations created by a pulse-pair.	 
Here, the anharmonic nonlinear source term $\propto \delta|\Delta|\Delta\rho^\mathrm{A,B}_2$ generates rephasing photon echo signals at $(-\omega_\mathrm{B}+\omega_\mathrm{A})+\omega_\mathrm{A}=(\omega_\mathrm{H},-\omega_\mathrm{H})$ 
and $(-\omega_\mathrm{A}+\omega_\mathrm{B})+\omega_\mathrm{B}=(\omega_\mathrm{H},2\omega_\mathrm{H})$ as well as pump-probe (Fig. 2H) and non-rephasing signals (Fig. 2I) at $(\omega_\mathrm{H},\omega_\mathrm{H})$ and $(\omega_\mathrm{H},0)$. 



In summary, we discover quantum echo signals arising from transient, yet sufficient long-lived, Higgs coherence in superconductivity. The Higgs echo signals are characterized by a delayed build-up time deviating from symmetric formation and a negative time signal from anharmonicity. These features set it apart from any other known echo phenomena.
Importantly, our observations cannot be solely explained by either Higgs or QP excitation alone. Quantum interference and anharmonicity between two phase-locked Higgs excitations, and their interactions with QP coherence, however, comprehensively explain all of our observations. 
 
\bibliographystyle{Science}

   
\noindent {\bf Funding:} THz quantum dynamics study was supported by the U.S. Department of Energy, Office of Basic Energy Science, Division of Materials Sciences and Engineering (Ames National Laboratory is operated for the U.S. Department of Energy by Iowa State University under Contract No. DE-AC02-07CH11358). 
Equilibrium characterization of superconductors and modeling were supported by the U.S. Department of Energy, Office of Science, National Quantum Information Science Research Centers, Superconducting Quantum Materials and Systems Center (SQMS) under the contract No. DE-AC02-07CH11359

\clearpage


\end{document}